INTERNATIONAL JOURNAL OF
# Biological Macromolecules

# *N,N,N*-Trimethyl chitosan as a permeation enhancer for inhalation drug delivery: interaction with a model pulmonary surfactant

**Jana Szabová[1,2]\*, Filip Mravec[2], Mostafa Mokhtari[3], Rémi Le Borgne[4], Michal Kalina[2] and Jean-François Berret[1]\***

[1]*Université Paris Cité, CNRS, Matière et systèmes complexes, 75013 Paris, France*
[2]*Materials Research Centre, Faculty of Chemistry, Brno University of Technology, Brno, Czech Republic*
[3]*Neonatal Intensive Care Unit, Hôpitaux Universitaires Paris – Saclay, Hôpital Universitaire de Bicêtre, and Espace Ethique/Île-deFrance, Hôpital Universitaire Saint-Louis - APHP , Paris, France*
[4]*Université de Paris, CNRS, Institute Jacques Monod, 75013 Paris, France*

**Abstract:** *N,N,N*-Trimethyl chitosan (TMC), a biocompatible and biodegradable derivative of chitosan, is currently used as a permeation enhancer to increase the translocation of drugs to the bloodstream in the lungs. This article discusses the effect of TMC on a mimetic pulmonary surfactant, Curosurf®, a low-viscosity lipid formulation administered to preterm infants with acute respiratory distress syndrome. Curosurf® exhibits a strong interaction with TMC, resulting in the formation of aggregates at electrostatic charge stoichiometry. At nanoscale, Curosurf® undergoes a profound reorganization of its lipid vesicles in terms of size and lamellarity. The initial micron-sized vesicles (average size 4.8 μm) give way to a froth-like network of unilamellar vesicles about 300 nm in size. Under such conditions, neutralization of the cationic charges by pulmonary surfactant may inhibit TMC permeation enhancer capacity, especially as electrostatic charge complexation is found at low TMC content. The permeation properties of pulmonary surfactant-neutralized TMC should then be evaluated for its applicability as a permeation enhancer for inhalation in the alveolar region.

## Highlights

- Pulmonary surfactant substitute Curosurf® is composed of lipid vesicles
- Permeation enhancer *N,N,N*-Trimethyl chitosan interacts with Curosurf® by charge complexation
- Interaction is maximum at electrostatic charge stoichiometry
- Synthetic cationic polymers show similar behavior to *N,N,N*-Trimethyl chitosan
- Interaction leads to the formation of micron-sized vesicle aggregates
- At nanoscale Curosurf® vesicles undergoes a profound change in size and lamellarity
- Complexation causes the immobilization of *N,N,N*-Trimethyl chitosan in aggregates

## Introduction

Inhalation is a powerful administration route for the local treatment of pulmonary diseases including asthma, chronic obstructive pulmonary disease, cystic fibrosis, lung cancer, or COVID 19 associated pulmonary embolism [1-3]. The lungs may also provide a route of administration to treat non-pulmonary diseases such as diabetes, Parkinson's disease and migraine [3, 4]. Due to the large surface area ($\sim$ 100 m$^2$) and thin air-blood interface in the alveolar region (where the gas exchange takes place), the absorption efficiency is high and the onset of action is rapid [5]. The low metabolic activity in the lungs and the avoidance of the first-pass effect make inhalation also suitable for peptide or protein-based drugs [1, 3]. However, due to their sizes, macromolecules and more generally drug nanocarriers permeate poorly across epithelial barriers [6, 7]. The alveolar epithelium is more permeable than other epithelia but still presents a kinetically restricted diffusive absorption for macromolecules, especially for those with $M_w$ > 1 kDa [8]. This issue could be overcome by permeation enhancers such as *N,N,N*-trimethyl chitosan polymers, hereafter abbreviated as trimethyl chitosan or TMC.

Studies have shown that permeation enhancers can increase local adsorption and translocation to the blood capillaries for macromolecules or active pharmaceutical substances that otherwise permeate poorly across the alveolar epithelium. Trimethyl chitosan is a biocompatible and biodegradable derivative of chitosan with three methyl groups on nitrogen which provide positive charges and water-solubility in a wide pH-range when the degree of quaternization is higher than 40% [9-12]. The permanent positive charges also ensure unique features such as muco-adhesion [10], reversible opening of tight junctions between epithelial cells [13], and antimicrobial effect *via* interaction with bacteria cell walls [14]. Tight junctions are depicted as highly hydrated complexes with fixed negative sites. At the cell level, changes in electrostatic binding mediated by TMC polymers could result in the alteration of the junction structure. This leads to their opening for biologically active compounds (e.g. for mannitol, buserelin, octreotide, ofloxacin or insulin [15, 16]) which are then transported by paracellular way [16]. TMC has also been studied as a drug and gene delivery system by different modes of administration including peroral, ocular, nasal, buccal, intravenous, pulmonary and rectal [9, 17].

The alveolar epithelium is composed of type I and type II alveolar cells: alveolar type I cells are involved in the oxygen/carbon dioxide gas exchange, whereas alveolar type II cells are responsible for the synthesis and secretion of pulmonary surfactant. These epithelial cells are connected through tight junctions [18]. The pulmonary surfactant consists of a thin (< 1 μm) layer at the surface of the alveolar epithelium and its role is to reduce the surface tension with air [19], prevent the alveoli collapse during exhalation, and facilitate their expansion during breathing [20, 21]. Pulmonary surfactant is formed by 90 % of lipids and 10 % of surfactant proteins, while lipids at the concentration of 40 g L$^{-1}$ are arranged into multilamellar vesicles [22-24]. Surfactant proteins SP-A and SP-D are large hydrophilic glycoproteins that bind pathogens and inhaled particles to facilitate phagocytosis. SP-B and SP-C are small hydrophobic proteins for regulating and maintaining surfactant interfacial properties [19, 25]. For drug nanocarriers reaching the alveolar space, pulmonary surfactant may represent a barrier with which they will interact, and potentially modify their therapeutic properties.

The interaction between polymers and native and mimetic pulmonary surfactants was studied in several publications [26-33]. The objective was to develop drugs to treat acute respiratory distress syndrome (ARDS) in adults, which results from the inactivation of the surfactant by an increased amount of proteins, e.g. such as albumin, hemoglobin, fibrinogen and fibrin in their composition. These proteins are adsorbed at the air/liquid interface and made the interface poorly accessible to the pulmonary surfactant. Hydrophilic polymers such as poly(ethylene glycol) (PEG) [27, 28, 30], dextran [27, 28], poly(vinylpyrrolidone) [28], hyaluronic acid (HA) [28, 31-33] and chitosan [29], as well as a combination of PEG or dextran mixture with HA [33] or with SP-B or SP-C [20] were found to reverse the effect of proteins and restored pulmonary surfactant function.

Beyond polymers, recent studies have been conducted on the interaction of model nanoparticles with the exogenous pulmonary surfactant [25], in particular the drug developed by Chiesi (Parma, Italy), Curosurf® [34-36]. It was found that cationic particles such as aluminum oxide ($Al_2O_3$) and amine-coated silicon dioxide ($SiO_2$) exhibit strong electrostatic interaction with the lipid membranes, leading to spontaneous nanoparticle-vesicle aggregation. Besides, living cell assays performed with Curosurf®-treated $SiO_2$ revealed that surfactant decreases the nanoparticle uptake by two orders of magnitude in A549 and NCI-H441 alveolar epithelial cells, irrespective of immersed or air-liquid interface cultures [35, 36]. This work supports the idea that the pulmonary surfactant plays a protective role against inhaled particles, as it slows down or mitigates their translocation to the blood stream. The effect of surfactant must therefore be taken into account in the predictive evaluation of TMC-based drug nanocarriers with enhanced permeation attributes.

The aim of this work is to study the effect of trimethyl chitosan polymer on the mimetic pulmonary surfactant Curosurf® under alveolar physiological conditions. Curosurf® is a low-viscosity lipid vesicle formulation containing 1% of the membrane proteins SP-B and SP-C [19]. It is administered to preterm infants with ARDS, and is nowadays recognized as a reliable surfactant model regarding its composition, structure and stability. The main result that emerges from this work is that trimethyl chitosan exhibits a strong interaction with Curosurf®. This interaction originates from electrostatic complexation between opposite charges, positive for TMC and negative for Curosurf®. This interaction furthermore leads to a profound reorganization of the lipid aggregates, in terms of size and lamellarity (*i.e.* the number of consecutive lipid bilayers within one vesicle [37]): The micron-sized multilamellar vesicles characteristic of Curosurf® are transformed into a connected network of smaller unilamellar vesicles upon TMC addition.

# 2 - Material and Methods

## 2.1 - Materials

*Polymers.* Medium molecular weight *N,N,N*-trimethyl chitosan (CAS 52349-26-5) with a degree of quaternization of 40-60 % quantified by [1]H-NMR and a zeta potential $\zeta$ = 44 ± 5 mV was purchased from Sigma Aldrich (Czech Republic). The molecular weight ($M_w$ = 172.4 kDa) and molar mass dispersity (Đ = 1.7) of the TMC macromolecules was determined by size exclusion chromatography (SEC) system (Infinity 1260, Agilent Technologies, USA) equipped with PL Aquagel-OH MIXED-H column (300x75 mm) coupled with multi-angle light scattering detection (MALS, Dawn Heleos II, Wyatt Technology, USA) and differential refractometry (dRI, Optilab T-rEX, Wyatt Technology, USA). Poly(diallyldimethyl ammonium chloride) (PDADMAC) of molecular weight $M_w$ = 13.4 kDa and $M_w$ = 26.8 kDa was purchased from Polyscience and from Aldrich. Poly(sodium 4-styrene sulfonate) (PSS) of molecular weight $M_w$ = 8.0 kDa was obtained from SRA Instruments. Their molecular weights and molar mass dispersity Đ were obtained from light-scattering detection (MALS, Dawn Heleos II, Wyatt Technology, USA) combined with SEC using a Novema (PDADMAC) and a Shodex (PSS) column [38]. The polymer characteristics are summarized in **Table I**, and their chemical formulae provided in **Figs. 1a-c**.

**Polymer characterization**

| Polymer | Provider | $M_n$ (kDa) | $M_w$ (kDa) | Đ | Charge |
|---------|----------|-------------|-------------|---|--------|
| PSS | SRA Instruments | 6.7 | 8.0 | 1.2 | − |
| TMC | Sigma Aldrich | 105.7 | 172.4 | 1.7 | + |
| PDADMAC | Polyscience Europe | 4.9 | 13.4 | 2.7 | + |
|  | Sigma Aldrich | 7.6 | 26.8 | 3.5 | + |

**Table I** : Characterization of the polymers used in this work. PSS, TMC and PDADMAC stands for poly(sodium 4-styrene sulfonate), *N,N,N*-trimethyl chitosan and poly(diallyldimethyl ammonium chloride), respectively. $M_n$ and $M_w$ denote the number- and weight averaged molecular weights, and Đ the molar mass dispersity.

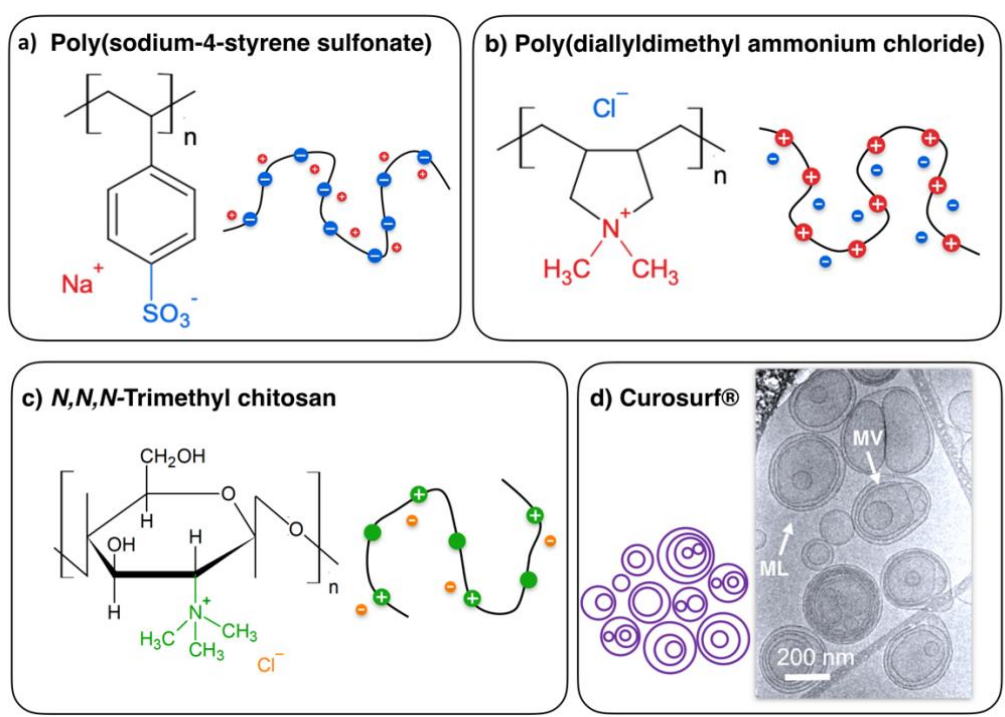

***Figure 1:*** *Chemical formulae and illustration of the polymers and vesicles used in this work.* ***a)*** *Poly(sodium-4-styrensulfonate) (PSS);* ***b)*** *Poly(diallydimethylammonium chloride) (PDADMAC);* ***c)*** *N,N,N-trimethyl chitosan (TMC) and* ***d)*** *Curosurf® lipid vesicles. The image in Fig. 1d displays multilamellar and multivesicular vesicles taken by cryo-transmission electron microscopy [34, 35].*

*Pulmonary surfactant.* Curosurf® (*Chiesi Pharmaceuticals*, Parma, Italy) is a porcine minced lung extract developed for the treatment of premature newborns with ARDS [39]. It is a 90:10 dispersion of phospholipids, including phosphatidylcholine, sphingomyelin, phosphatidylglycerol and of the surfactant proteins SP-B and SP-C, at a total concentration of 80 g L$^{-1}$ [34, 40]. According to a recent study, premature infants with neonatal respiratory distress syndrome have an almost 20% better chance of survival when treated with Curosurf® compared to those who receive the other surfactant therapies [41]. Curosurf® composition and its comparison with native human surfactant can be found the **Supplementary Information S1**. The pH of Curosurf® is adjusted by the provider to an average value of pH 6.3 by addition sodium bicarbonate, varying between 5.5 and 6.5 and depending on the batches [40, 42]. Concerning the stability of Curosurf® samples in the long term, these dispersions are stable over several months and years. The same applies to mixed polymer-vesicle aggregates after their formation. **Fig. 1d** displays a representative cryo-TEM image of Curosurf® structure determined at 5 g L$^{-1}$, and reveals the presence of multilamellar and multivesicular vesicles. Data collected using cryo-TEM, dynamic light scattering, nanoparticle tracking analysis and optical microscopy have shown that Curosurf® dispersions exhibit a wide distribution of multilamellar and multivesicular vesicle sizes from 50 nm and a few micrometers, the smallest being the most abundant. Curosurf® samples were kindly provided by Dr. Mostafa Mokhtari and his team from the neonatal service at Hospital Kremlin-Bicêtre, Val-de-Marne, France.

## 2.2 - Extruded Curosurf®

To control the vesicular size, extrusion of dilute Curosurf® dispersions was performed using an Avanti Mini Extruder (Avanti Polar Lipids, Inc. Alabama, USA). The initial 80 g L$^{-1}$ stock sample was first diluted with DI-water to 1 g L$^{-1}$ and extruded 20 times through Whatman Nucleopore polycarbonate membranes (pore size 200 nm) at the temperature of 37 °C. This technique leads to a narrow distribution of vesicles centered around 180 nm, in good agreement with earlier reports [34, 40].

## 2.3 - Mixing protocols and Job scattering plots

The interaction between the oppositely charged macromolecules and/or macromolecules and vesicles was investigated by the continuous variation method [43, 44]. This method was originally developed one century ago by P. Job in the context of ionic association [43, 44] and later adapted for a wide range of techniques, including for light and small-angle neutron/X-ray scattering [34, 40, 45]. Solutions were prepared in DI-water in the same conditions of pH (pH 6.3) and concentration ($c$ = 1 g L$^{-1}$). The pH was adjusted by titration with nitric acid or ammonium hydroxide when necessary. The PSS, PDADMAC and TMC dispersions were filtered (cellulose acetate filter, 0.45 μm) and assessed by light scattering. Oppositely charged polymers (PSS/PDADMAC, PSS/TMC) and oppositely charged vesicles and polymers (Curosurf®/PDADMAC, Curosurf®/TMC) were then formulated at the different volume ratio $X$ between 10$^{-3}$ and 10$^3$. Series of 12-14 mixed samples were made for each pair of oppositely charged species, the dispersions having a total organic content of 1 g L$^{-1}$. For oppositely charged vesicles and cationic polymers, $X$ was defined as $c_{Curo}/c_{Pol}$, where $c_{Curo}$ and $c_{Pol}$ denotes the Curosurf® and polymer weight concentration, respectively. After mixing, the dispersions were rapidly stirred and let 10 minutes to equilibrate. Then the static scattering intensity $I_S$ and hydrodynamic diameter $D_H$ were measured in triplicate at room temperature (T = 25 °C). For non-interacting species, e.g. for same charge polymers, the scattering intensity is the sum of the $X$ = 0 and $X$ = ∞ intensities, weighted by their actual concentrations, leading $I_S^0(X) = \left(I_S(X=0) + XI_S(X=\infty)\right)/(1+X)$. The light scattering data were systematically compared to this reference to ensure the occurrence of interaction between species. All studied formulations were prepared in duplicate and light scattering results were averaged over the two series of samples obtained. For the oppositely charged polymers, $X$ was translated into the charge ratio $Z_{-/+} = [-]/[+]$, where $[-]$ and $[+]$ denotes the molar concentrations of negative and positive charged carried by the chains. In this case, the continuous variation method works as a titration experiment and allows the determination of the structural charges (and not the effective charge) carried by the colloidal species [46], as well as the charge stoichiometry. In the sequel of the paper, the scattering intensity or hydrodynamic diameter as a function of $X$ or $Z$ are termed Job scattering plots or diagrams.

## 2.4 - Membrane labeling

PKH67 (Sigma Aldrich) is an amphiphilic fluorescence probe, where two aliphatic chains are inserted in the bilayer and headgroups are emitting fluorescence in the green (emission maximum at 502 nm). This probe is usually used to stain living cells, but it was alternated for vesicle labeling [31,35]. The stock solution of Curosurf® (2 g L$^{-1}$) and PKH67 (2×10$^{-6}$ M) was prepared by dilution of deionized water of their stock solutions. These solutions were combined in equal volume, vortexed for 10 seconds, and let in the dark for 15 minutes to ensure the insertion of the probe into the bilayer of vesicles.

## 2.5 - Static and dynamic light scattering

Scattered light intensity $I_S$ and hydrodynamic diameter $D_H$ were performed on a ZetaSizer NanoZS (Malvern Panalytical). With the Malvern software, $I_S$ corresponds to the derived count rate, expressed in kcps. The He-Ne laser (633 nm) was used to illuminate the sample and the scattered light was collected in backscatter detection mode at an angle of 173°. The second-order autocorrelation function was analyzed using the cumulative and CONTIN algorithms to define the average diffusion coefficient $D_0$ of individual particles. The $D_H$ was then calculated according to the Stokes-Einstein relation $D_H = k_B T/3\pi\eta D_0$ where $k_B$ is the Boltzmann constant, $T$ is the temperature, and $\eta$ is the solvent viscosity. In all results, $D_H$ corresponds to the $Z_{Ave}$ value obtained from the cumulant analysis. Samples were measured in triplicate at the temperature of 25 °C after 60 seconds of equilibration.

## 2.6 - Optical and Fluorescence microscopy

Images were acquired on an IX73 inverted microscope (Olympus) equipped with a 40× objective. An Exi Blue camera (Q-Imaging) and Metaview software (Universal Imaging Inc.) were used as the acquisition system. The illumination system "Illuminator XCite Microscope" produced a white light, filtered for observing a green signal in fluorescence (excitation filter at 470 nm – bandwidth 40 nm and emission filter at 525 nm – bandwidth 50 nm). Samples of Curosurf® for optical microscopy were prepared as described above and diluted 100 times for background reduction. These samples were placed on the glass plate that was previously dip-coated with PDADMAC$_{26.8k}$ to increase adsorption and sealed into a Gene Frame dual-adhesive system (Abgene/Advanced Biotech). The labeled vesicles of Curosurf® (1 g L$^{-1}$) were combined with different polycations to observe the result of interaction between these two components. The concentration ratios $X = c_{Curo}/c_{Pol}$ were chosen according to the results obtained from the Job scattering plot. The concentration ratio $X$ selected for TMC was $X = 10$, for PDADMAC$_{13.4k}$ was $X = 30$, and for PDADMAC$_{26.8k}$ was $X = 100$. Then, the preparation of the samples was done as for Curosurf® alone, but the surface of the glass plates was not modified. Images were digitized and treated by the ImageJ software and plugins (http://rsbweb.nih.gov/ij/).

## 2.7 - Transmission electron microscopy

Curosurf® (80 g L$^{-1}$) or Curosurf® with TMC (combined in the concentration ratio $X = 10$) was centrifuged (30 minutes, 12 000 rpm) to form pellets. The dispersion was fixed with 2.5 wt. % glutaraldehyde and 4 wt. % PFA in PBS buffer. The sample was stored in the fixative for 3 h at room temperature and then kept at 4 °C. Afterwards, the sample was centrifuged at 500 – 1000 g for 3 min to obtain a pellet. After several washing steps with buffer, the sample was subsequently post-fixed in 1% osmium tetroxide, and then pelleted in 1% agarose gel. The sample in agarose was cut in small pieces of less than 1mm³, dehydrated in an ascending ethanol series and embedded in Epon (Agar low viscosity resin) and polymerized overnight at 60 °C. Ultrathin 70-nm sections were cut using an ultramicrotome UC6 (Leica). The sections were post-stained with uranyl acetate and lead nitrate, and analyzed with a 120-kV transmission electron microscope Tecnai 12 (ThermoFisher Scientific) using the 4-k camera OneView and the GMS3 software (Gatan).

# 3 - Results and discussion

### 3.1 - Electrostatic properties of *N,N,N*-trimethylchitosan

The electrostatic charge density of trimethyl chitosan at pH 6.3 was evaluated using the continuous variation method [38, 43, 44]. As shown in previous studies [34, 45, 47], complexation between oppositely charged macromolecules gives rise to a scattering peak when the intensity is plotted as a function of the charge ratio $Z$, and the maximum is centered on the charge stoichiometry ($Z = 1$). To evaluate the charge density of TMC, we applied this method using poly(sodium 4-styrene sulfonate) as a complexing agent. **Fig. 2a** displays the scattering intensity $I_S(Z_{-/+})$ of mixed PSS/TMC solutions versus $Z_{-/+}$. There, $I_S(Z_{-/+})$ exhibits a sharp maximum resulting from the formation of electrostatic complex and/or coacervation droplet formation [48-50]. The peak position in the Job scattering diagram is also associated with the neutralization of the electrostatic charges [38]. The maximum charge ratio for this diagram is located at the critical level $Z_C = 0.4$, indicating that 40% of positively charged TMC monomers are accessible to interact with the negatively charged PSS. The second local maximum observed at $Z = 4$ in **Fig. 2a** could be attributed to the primary and secondary amino groups present along the polymer backbone following TMC quaternization (**Fig. 2b**). As previously reported, fewer than three methyl groups on nitrogen result in a pH-dependent charge and potential insolubility under physiological conditions [11, 51, 52]. Our working pH being close to physiological conditions, chitosan and dimethyl chitosan are not fully soluble, but some of the amino groups could be protonated which can result in interaction with negatively charge PSS. The occurrence of a second maximum could indicate a second titration, here for primary and secondary amines which would not have titrated during the first process [10]. Overall, the data show a marked electrostatic interaction of TMC with PSS macromolecules, a result that corroborates earlier reports of TMC interacting with alginates [53, 54], dextran and hyaluronate [55] or different anionic compounds such as Aβ40 peptides [56], solid lipid nanoparticles [57], microemulsions [58], niosomes [59] or liposomes [60-66].

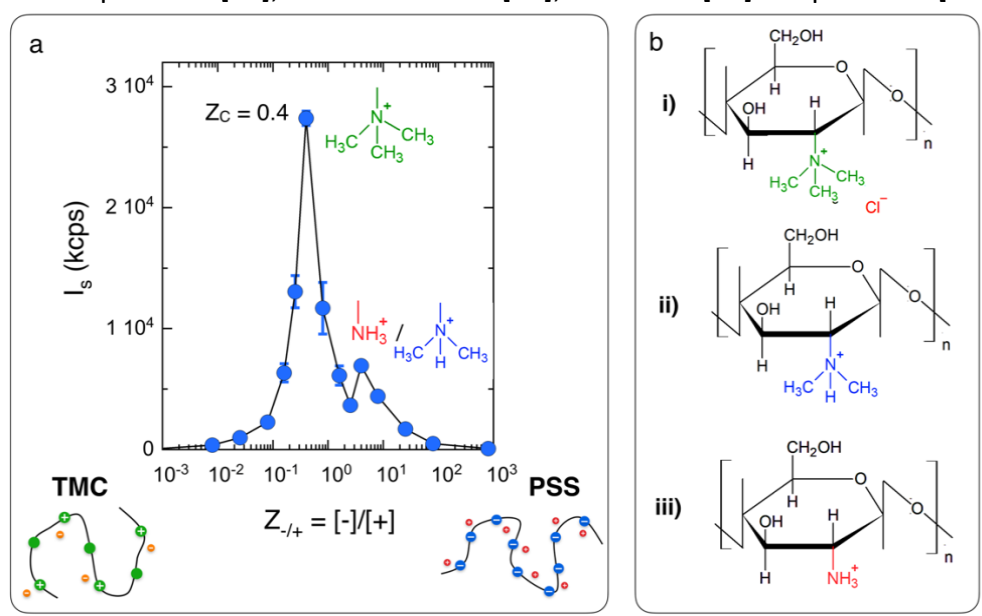

***Figure 2: a)*** *Job scattering plot $I_S(Z_{-/+})$ of mixed polymers dispersions made from N,N,N-trimethyl chitosan (TMC) and poly(sodium-4-styrenesulfonate) at pH 6.3, where $Z_{-/+}$ denotes the charge ratios. The maximum is located at $Z_{-/+} = 0.4$ and a second local maximum is found around $Z_{-/+} = 4$. Following the approach quoted in Ref. [38], the position of the main peak allows to derive the percentage of charged TMC monomers, here estimated at 40%.* ***b)*** *Possible functional groups in the polymer chain: I)*

*N,N,N-trimethyl chitosan, II) N,N-dimethyl chitosan III) chitosan. This latter corresponds to the original unmodified chitosan macromolecule. Their structures are illustrated at the physiological pH.*

### 3.2 - Interaction between polycations and Curosurf®

We now turn to the electrostatic complexation between oppositely charged trimethyl chitosan and Curosurf® vesicles. As in the previous section, the continuous variation method is applied using here a combination of static and dynamic light scattering. For comparison, the complexation scheme is also assessed using PDADMAC polymers of lower molecular weight than TMC ($M_w$ = 172.4 kDa, Đ = 2.7) [38]. For this later, two values of molecular weights ($M_w$ = 13.4 kDa, Đ = 2.7 and 26.8 kDa, Đ = 3.5) are investigated. This comparison allows us to evaluate the effect of charge density and polycation molecular weight on the interaction with the lipidic membrane. In **Fig. 3**, Job scattering plots of the intensity and hydrodynamic diameter of polycations and Curosurf® are presented. There, the mixing ratio $X = 10^{-3}$ corresponds to the sample containing only polymer (at 1 g L$^{-1}$) and that of $X = 10^3$ to the sample containing only Curosurf® (at 1 g L$^{-1}$). This choice is illustrated in the figure by schematics of polymer chains and of vesicles. The continuous lines in green for the intensity (**Figs. 3a**, **3d** and **3g**) result from the assumption that polymers and vesicles are not interacting, as discussed in **Section 2.4**. The data in **Fig. 3** reveals that the polycations interact strongly with Curosurf®. In the Job scattering diagrams, $I_S(X)$ systematically exceed the predictions for non-interacting species (**Figs. 3a, 3d and 3g**). Moreover, a peak in $D_H(X)$ is observed at high Curosurf® concentrations ($X$ = 10-100) revealing micron-sized particles, *i.e.* sizes much larger than the hydrodynamic diameters of pristine dispersions (**Figs. 3b, 3e and 3h**). Finally, the change in turbidity of the mixed dispersions (**Figs. 3c, 3f and 3i**) is further evidence of the formation of large polymer/vesicle structures. Of note, the critical mixing ratios associated with the $I_S(X)$ and $D_H(X)$ peaks are different, a result that is attributed to the turbidity of the solutions that decreases the scattering intensity for $X \gg 1$. For this reason, the critical ratios $X_C$ are determined from the $D_H$-maxima. We found here $X_C$ = 20 ± 5 for TMC, 30 ± 7 for PDADMAC$_{13.4k}$ and $X_C$ =100 ± 20 for PDADMAC$_{26.8k}$ (**Table II**).

For the past decade, the continuous variation method has been applied to various colloidal systems, including macromolecules, amphiphiles and nanoparticles [67, 68]. Overall, mixed polymer/amphiphiles and polymer/nanoparticle dispersions have been studied in detail, and disclose similar scattering patterns in the Job representation, with critical values of $X_C$ around 1. For the polymer/vesicle systems studied here, however, we find maxima in the region $X$ = 10-100. As shown below, this result can be explained quantitatively in the context of processes related to electrostatic charge pairing and co-assembly. Based on the lipid composition of the endogenous pulmonary surfactant (**Supplementary Information S1**), around 10% of the lipids (e.g. phosphatidylinositol and phosphatidylglycerol) are negatively charged, leading to a surface charge density of $\sigma_S$ = -0.17$e$ nm$^{-2}$ [21, 35]. Regarding Curosurf®, electrophoretic measurements performed on native and extruded samples have revealed zeta potential values of $\zeta_{native}$ = -54 mV and $\zeta_{extruded}$ = -24 mV at the lung physiological conditions [40]. These results confirm that surfactant membranes are on average negatively charged. With this assumption, it is possible to calculate the relationship between the mixing ratio $X = c_{Curo}(X)/c_{Pol}(X)$ and the charge ration $Z_{-/+} = (-)/(+)$, assuming that the TMC chains are only charged to 40% and the PDADMACs to 100% under the current physico-chemical conditions. Doing so, we find that the critical values of the $Z_C$ are 0.90 ± 0.22, 0.34 ± 0.08 and 0.90 ± 0.22 for the TMC, and the two PDADMAC of different molecular weights, respectively (**Table II**). Except for PDADMAC$_{13.4k}$, the values are close to unity, and suggest that electrostatic complexation with charge pairing and association between the monomers and lipidic membranes takes place here. The shift of $X_C$ to higher values in the Job scattering plots (compared to what was found for particles and polymers) can thus be understood as a combination of two effects, the high molecular weights of lipids compared to that of the polyelectrolyte monomers, and the small proportion of lipids that are actually charged. This phenomenon highlights that only a minute amount of cationic polymers is needed to reach charge neutrality, and induce structural changes in the pulmonary surfactant.

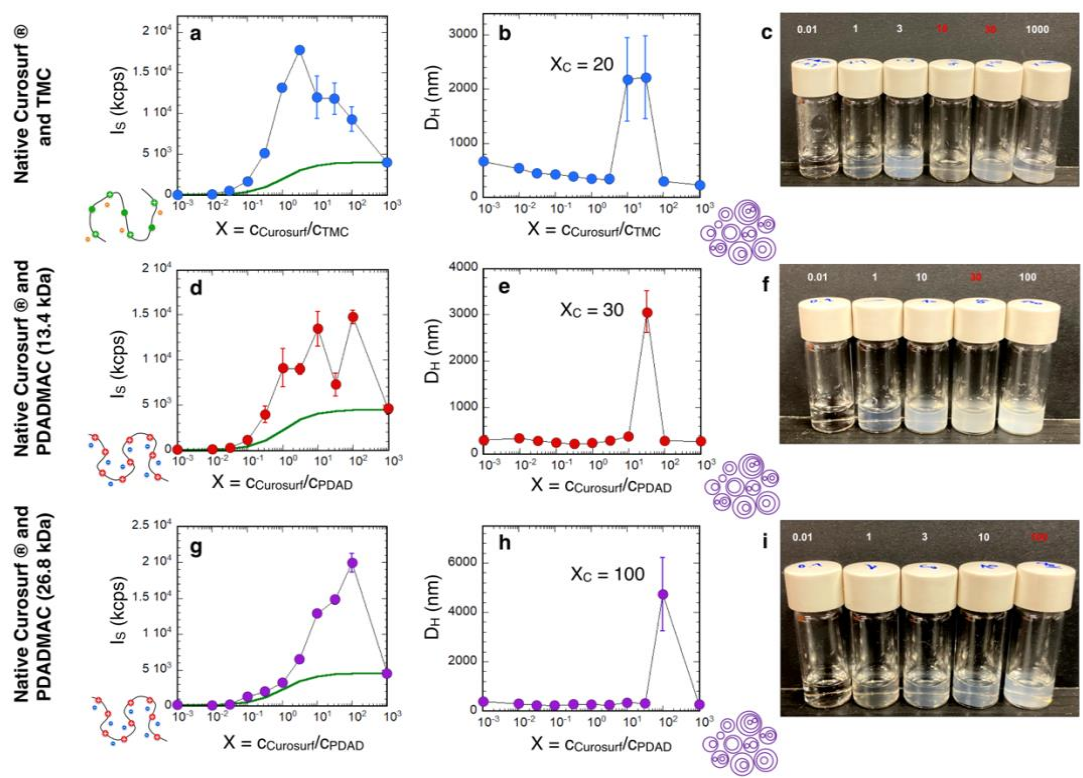

***Figure 3**: **a, b)** Job scattering plots displaying the scattering intensity and the hydrodynamic diameter versus $X$ of native Curosurf® and N,N,N-trimethyl chitosan (TMC) mixed dispersions. There, the continuous line in green indicates non-interaction case between mixed components. **c)** Image of TMC/native Curosurf® vials. The labels indicate the $X$-values and the one in red refers to the critical mixing ratio $X_C$ determined from the $D_H(X)$-peak. **d, e, f)** Same as **a)**, **b)** and **c)** for poly(diallyldimethylammonium chloride) (PDADMAC) of molecular weight 13.4 kDa and native Curosurf®. **g, h, i)** Same as **a)**, **b)** and **c)** for PDADMAC of molecular weight 26.8 kDa and native Curosurf®.*

## 3.3 – Visualization of micron-sized polymer/vesicles aggregates by optical microscopy

Regarding electrostatic complexation, two main scenarios have been reported yet. Studies have shown either a thermodynamic phase transition leading to coacervation [48, 50], as for polymer/polymer or polymer/amphiphile systems, or the formation of large disordered aggregates, whose size, structure and charge depend on the charge ratio $Z$. To address the question of the phase induced by complexation, phase contrast and fluorescence optical microscopy have been performed on pure and mixed dispersions. **Fig. 4a and 4b** display Curosurf® vesicles observed in 40× phase-contrast microscopy and in fluorescence, respectively. The vesicles are identified as separated objects adsorbed at the glass substrate. With fluorescence, the vesicles appear as bright spots co-localized with those seen in phase-contrast [47]. An analysis of the merged image (**Fig. 4c**) led to the outcome that around 80% of the vesicles are labeled and that their structure is not modified by the staining. **Fig. 4d and 4e** show images of a representative aggregate obtained by mixing of TMC and native Curosurf® at the charge stoichiometry. This aggregate has a disordered shape and a size of about 100 µm, *i.e.* much larger than single vesicles. The labelling of the lipid bilayer via the fluorescence probe PKH67 enables us to show that the aggregates are fluorescent, and result from the clustering of numerous vesicles (**Fig. 4e**). The merge of the phase contrast and fluorescence images shows again an excellent colocalization (**Fig. 4f**), suggesting that this aggregate results from the electrostatic assembly between polymers and lipidic membrane. Overall, the data obtained with PDADMAC give identical results to those of the TMC (**Fig. 4g-l**). We can notice that the size of the PDADMAC$_{26.8k}$/native Curosurf® aggregates are smaller than in the previous cases. Furthermore, the optical microscopy results do not show significant effects of polyelectrolyte charging rate (40% for TMC and 100% for PDADMAC) or molecular weight and at the same time they rule out the hypothesis of a coacervation-type phase separation [69]. Additional images of the mixed polycation/vesicles are shown in **Supplementary information S2**.

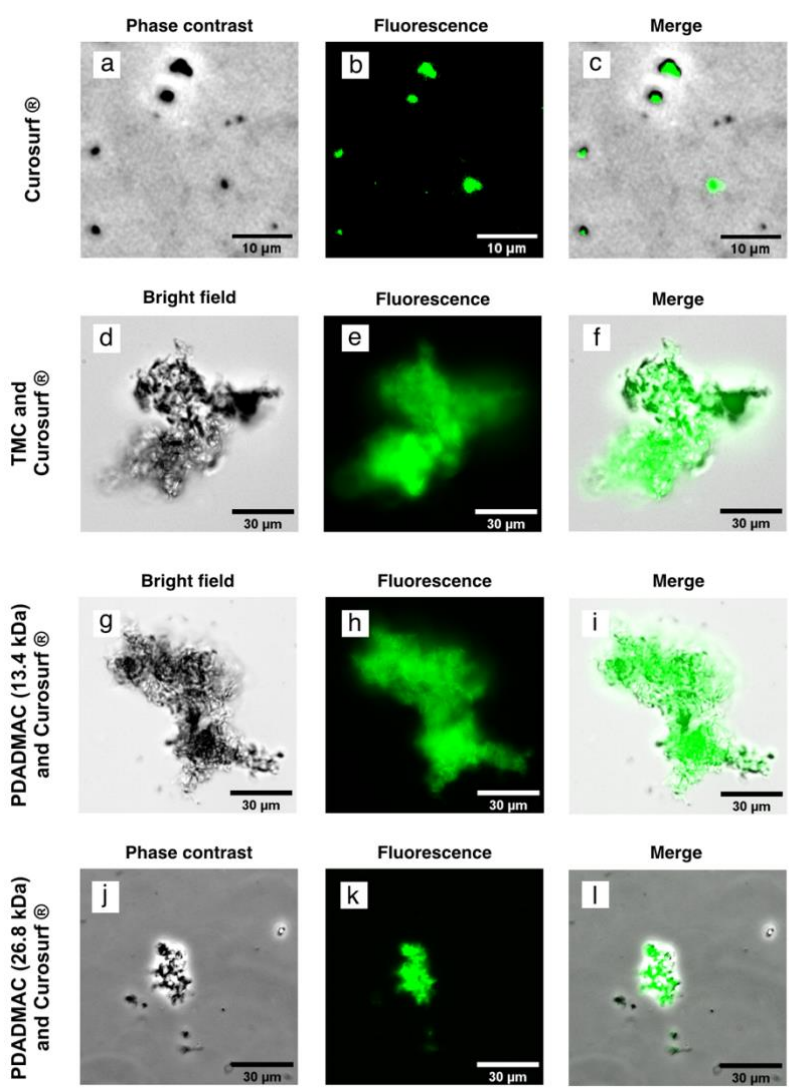

***Figure 4:*** *– **a,b)** Phase-contrast and fluorescence optical microscopy of Curosurf® vesicles (1 g L$^{-1}$) adsorbed on PDADMAC-coated glass substrate. **c)** Merge of the images in a) and b) showing colocalization. **d,e)** Bright-field optical microscopy of an aggregate obtained from N,N,N-trimethyl chitosan (TMC) and Curosurf® at charge stoichiometry. **f)** Merge of the images in d) and e) showing*

*colocalization. **g, h, i)** Same as d,e,f) for poly(diallydimethylammonium chloride) with a molecular weight of 13.4 kDa. **j, k, l)** Same as d,e,f) for poly(diallydimethylammonium chloride) with a molecular weight of 26.8 kDa.*

### 3.4 – Role of extrusion on the polycation/vesicle behavior

As mentioned earlier, the lipids present in the Curosurf® are assembled in the form of multilamellar and multivesicular vesicles of widely dispersed sizes, between 50 nm and 5 μm. In this section, we aimed to find out whether the vesicle structure and lamellarity [37] plays a role in the interactions with the polycations. To address this issue, dilute Curosurf® dispersions were extruded using 200 nm pore-size polycarbonate membranes [35]. As a result, dispersions of unilamellar vesicles of uniform size around 180 nm were obtained [35]. **Fig. 5** summarizes the results found by light scattering and optical microscopy with the three pairs of polycation/extruded Curosurf® samples, with the polymers being TMC and the two PDADMAC batches. Essentially, the $I_S(X)$ and $D_H(X)$ Job scattering diagrams, presented in **Fig. 5a-b, 5d-e and 5g-h** respectively are similar to those obtained for native Curosurf®. A strong scattering intensity is observed as well as marked peaks of the hydrodynamic diameter. We found here for critical ratio $X_C$ = 30 ± 7 for TMC, 30 ± 7 for PDADMAC$_{13.4k}$ and $X_C$ = 100 ± 20 for PDADMAC$_{26.8k}$, in good agreement with data on non-extruded Curosurf®. As for native dispersions, these values provide charge ratios close to 1, confirming the electrostatic character of the interaction between polymer and vesicles (**Table II**). It should be noted that the size of the aggregates in optical microscopy is smaller than that obtained with native Curosurf®, a result that can be attributed either to the vesicle sizes (200 nm) or to their lower zeta potential [40]. These results corroborate studies showing that the neutralization of liposomal charges with polymers leads to an effective polymer/liposome complexation, this phenomenon being associated with irreversible lipid rearrangements, polymer corona formation, or membrane fusion [70]. Additional images of mixed aggregates for the three system under scrutiny can be found in the **Supplementary Information S3**.

**Critical mixing ratios for polycation/Curosurf® mixed dispersion**

| Polymers | Vesicles | $X_C$ | $Z_C$ |
|---|---|---|---|
| TMC | Native Curosurf® | 20 ± 5 | 0.90 ± 0.22 |
| | Extruded Curosurf® | 30 ± 7 | 1.34 ± 0.31 |
| PDADMAC$_{13.4k}$ | Native Curosurf® | 30 ± 7 | 0.34 ± 0.08 |
| | Extruded Curosurf® | 30 ± 7 | 0.34 ± 0.08 |
| PDADMAC$_{26.8k}$ | Native Curosurf® | 100 ± 20 | 1.12 ± 0.22 |
| | Extruded Curosurf® | 100 ± 20 | 1.12 ± 0.22 |

**Table II:** *List of critical mixing ratios $X_C$ for polymer/Curosurf® samples at which the hydrodynamic diameter exhibits a maximum (**Fig. 3 and 5**). The charge ratio $Z_C$ is calculated assuming a charge density of -0.17 e nm$^{-2}$ for the Curosurf® membrane.*

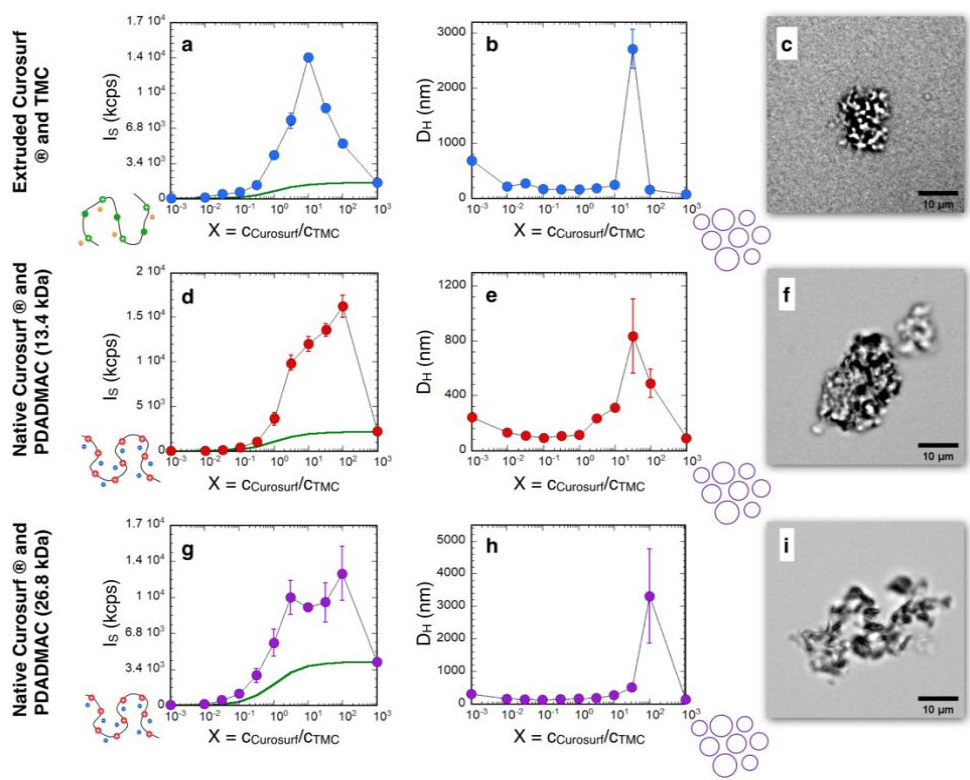

***Figure 5***: ***a, b)*** *Job scattering plots displaying the scattering intensity and the hydrodynamic diameter versus X of extruded Curosurf® and N,N,N-trimethyl chitosan (TMC). The continuous line in green indicates non-interaction case between mixed components.* ***c)*** *Bright-field optical microscopy of an aggregate obtained from TMC and extruded Curosurf® at charge stoichiometry, corresponding here to a mixing X = 30.* ***d, e, f)*** *Same as* ***a), b)*** *and* ***c)*** *for PDADMAC of molecular weight 13.4 kDa and extruded Curosurf®.* ***g, h, i)*** *Same as* ***a), b)*** *and* ***c)*** *for PDADMAC of molecular weight 26.8 kDa and extruded Curosurf®.*

### 3.5 – Structure of the polymer/Curosurf® aggregates monitored by transmission electron microscopy

When polyelectrolytes adsorb onto a surface of opposite charge, whether on a flat substrate or on a (nano)particle, chain conformations such as loops, trains and dangling ends of short polyelectrolyte segments are found, resulting in a nanometer-thin and highly charged layer [71]. When interaction occurs with self-assembled structures such as amphiphilic micelles or lipid membranes, the question arises whether the self-assembly retains its microstructure after interaction. Therefore, optical and transmission electron microscopy experiments were performed on 80 g L⁻¹ Curosurf® dispersions with and without TMC. Regarding TEM, the specimens were pre-fixed with glutaraldehyde and paraformaldehyde, embedded in resin and cut into a 70 nm thin slice prior to observation. For the TMC/Curosurf® mixed sample, the mixing ratio $X = 10$ was chosen in agreement with the position of the maximum hydrodynamic diameter in the Job scattering plot (**Fig. 3b**).

In **Fig. 6**, we compare the optical and electronic microscopy data on the two samples mentioned above. At the micrometer scale, optical microscopy shows no apparent change in vesicle size, dispersity or morphology with or without TMC (**Figs. 6a and 6b**). At the nanometer scale, the image in **Fig. 6c** first displays a 3.3 μm large multilamellar vesicle of native Curosurf®. The close-up view from **Figs. 6e** reveals more clearly the succession of interspaced lamellae at a characteristic distance of about 20-50 nm. The observed fluctuations in the membrane contours are linked to surface tension and/or elastic properties of the lipid bilayer [36]. The outcome in **Figs. 6c** and **6e** is representative of Curosurf® structure at clinical concentration (80 g L$^{-1}$), and in good agreement with earlier reports [34, 72, 73]. Following the addition of TMC, the lipid membranes are drastically modified. In the example shown in **Figs. 6d**, micron-sized multilamellar vesicles are no longer visible, but instead a multitude of smaller objects mainly unilamellar, that appear to form a complex network of interconnected vesicles. In the lower part of the figure, it is possible to observe a multilamellar vesicle, however of a smaller size compared to the native ones. These results bear some similarities with those obtained in the case of the addition of cationic Al$_2$O$_3$ nanoplatelets to Curosurf®, a result that was also accounted for by electrostatic-driven nanoplatelets adsorption to the membranes [74]. In the latter case, the alumina nanoplatelets do not carry amines on their surface but Al-OH$_2$$^+$ hydration complexes of high charge density (+7.3e nm$^{-2}$) [38, 75]. Additional optical and TEM images of Curosurf® and TMC/Curosurf® samples can be found in the **Supplementary Information S4 and S5**.

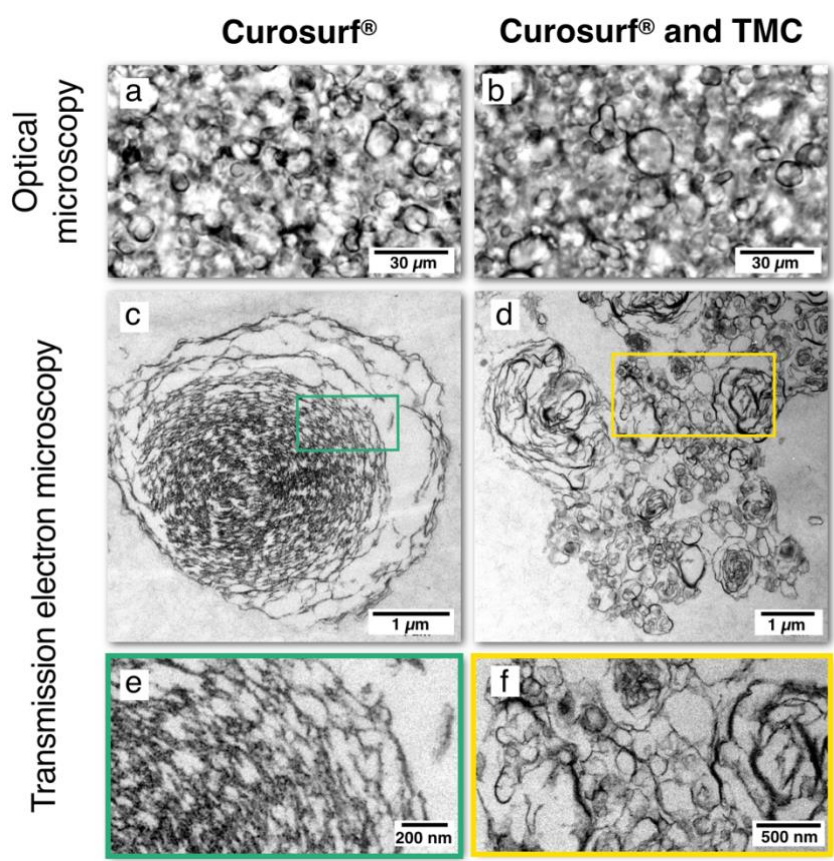

**Figure 6:** ***a, b)*** *Phase-contrast optical microscopy of Curosurf® (80 g L$^{-1}$) and mixed TMC/native Curosurf® at a concentration of 80 g L$^{-1}$ and a ratio X = $c_{Curo}/c_{Pol}$ = 10.* ***c, d)*** *Transmission electron microscopy of Curosurf® (80 g L$^{-1}$) and Curosurf® with TMC in the concentration ratio X = 10. e, f) Boxed areas showing a close-up view of Curosurf® (80 g L$^{-1}$) in c) and Curosurf® with TMC in the concentration ratio X = 10 in d).*

To gain a better understanding of the size reduction and lamellarity changes caused by TMC, an analysis of the size and dispersity of the vesicles was carried out. In this treatment, both multilamellar, multivesicular and unilamellar vesicles were counted. For native Curosurf®, the analysis was conducted on $n$ = 115 different objects comprised between 1 and 10 μm and showing mainly multilamellar or multivesicular structures. For this population, we found a broad distribution centered around 4.8 μm and dispersity $s$ = 0.41 (**Fig. 7a**). The dispersity $s$ is defined as the ratio between the standard deviation and the average size. An optical microscopy analysis of vesicle sizes on non-centrifuged Curosurf® performed by Thai et al. [76] also revealed a broad size distribution, centered on a median diameter of 3.3 µm, with a dispersity of 0.40 (**Supplementary Information SIXX**), in good agreement with the data in **Fig. 7a**. In the presence of TMC, 345 objects were examined, giving rise to a distribution with two maxima, one centered at 0.33 µm (with a dispersity $s$ = 0.45) and the other one at 2.4 µm ($s$ = 0.51), as displayed in **Fig. 7b**. As we anticipated, the population of the largest vesicles consists almost exclusively of multilamellar and multivesicular vesicles, whereas the population of the smallest vesicles is composed of unilamellar vesicles. The addition of TMC thus yields a reduction in size from 4.8 to 2.4 µm and in relative frequency from 0.25 to 0.03. This result suggests that the interaction with TMC led to a profound reorganization of the Curosurf® membranes: on the one hand, there is a decrease in proportion and size of the initial large vesicles; on the other hand, we observe the formation of a three-dimensional connected network of unilamellar vesicles, well identified from TEM data. Under these conditions, TMC present in topical lung drug formulations may lead to similar interaction with pulmonary surfactant, and inhibits its permeation enhancing properties.

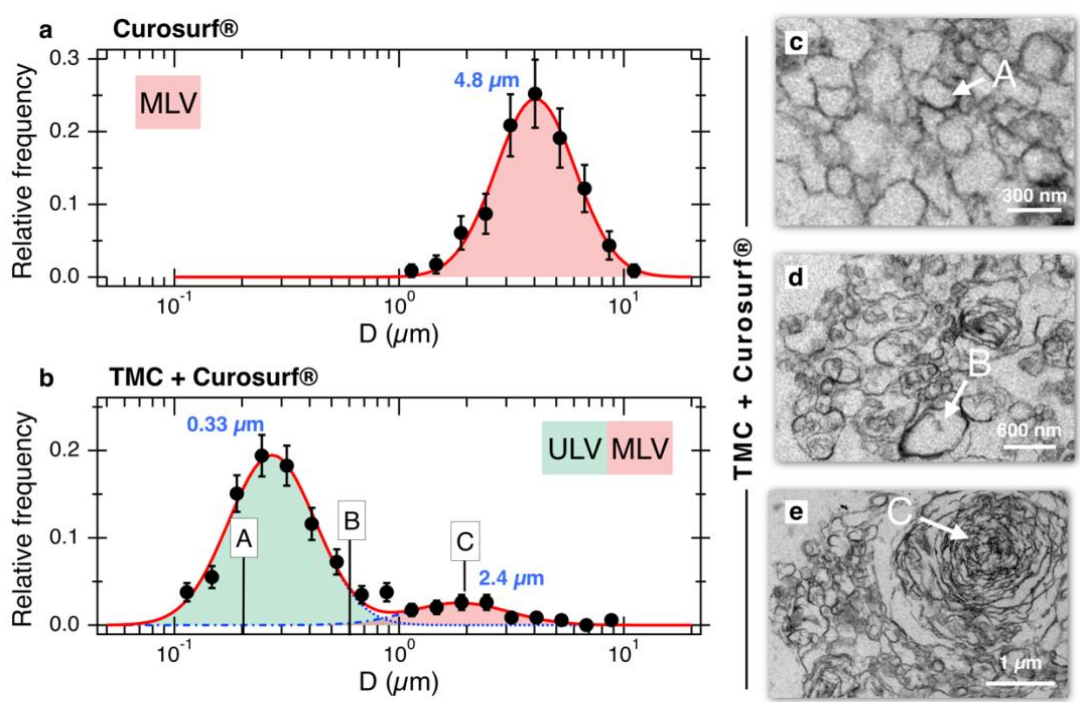

***Figure 7: a)*** *Size distribution of multilamellar and multivesicular objects obtained by transmission electron microscopy on a fixed native Curosurf® sample (80 g L⁻¹). The continuous curve in red outlines a vesicle population centered at 4.8 µm with a dispersity of 0.41.* ***b)*** *Size distribution of unilamellar and multilamellar vesicles obtained from mixed TMC/native Curosurf® sample at a concentration of 80 g L⁻¹ and a ratio $X = c_{Curo}/c_{Pol}$ = 10. The labels A, B and C refer to vesicle size shown in* ***Fig. 7c, 7d*** *and* ***7e*** *respectively.* ***c, d, e)*** *Illustration of TMC/native Curosurf® vesicles with characteristic sizes of 200 nm (A), 600 nm (B) and 2 µm (C).*

## Conclusion

In this study, the phase behavior of a pulmonary surfactant mimetic in the presence of *N,N,N*-trimethyl chitosan was examined. Charge titration experiments were first performed to determine the available positive charges along TMC chains. To this aim, narrow molecular weight dispersity poly(styrene sulfonate) was chosen as the model interaction partner, in conjunction with the continuous variation method. The results show that TMC has roughly 40% positive charges available for interaction with PSS. Subsequently, the interaction of *N,N,N*-trimethyl chitosan with the Curosurf®, a suitable model for pulmonary surfactant [36], was studied. Poly(diallyldimethyl ammonium chloride), a synthetic polycation widely used in formulations with two different molecular weights, was also studied with Curosurf® as a reference to evaluate interactions between oppositely charged polymers and lipid membranes. The Curosurf® model system was used for the study in native and extruded form. With both native and extruded Curosurf®, a strong interaction was noted, characterized in particular by the formation of aggregates and electrostatic charge compensation. Based on these observations, critical charge ratios were determined for TMC and PDADMAC polycations and Curosurf® in native and extruded form, confirming that the interactions between species were dominated by the charges, and by the complexation of the latter. These results are in good agreement with those of the literature describing charged polymer/liposome interaction [70]. The result of the interaction between TMC and Curosurf was further studied using Transmission Electron Microscopy. Upon TMC addition, transmission electron microscopy on fixed samples shows that Curosurf® undergoes a profound reorganization of the lipid vesicles in terms of size and lamellarity. The initial micron-sized multilamellar vesicles (average size 4.8 µm) give way to a foam-like network of connected unilamellar vesicles about 300 nm in size. Under such conditions, it is anticipated that the neutralization of TMC cationic charges by the pulmonary surfactant will lead to the de-activation of its permeation enhancer capacity, especially as the charge compensation is observed at low TMC content. Permeation properties of TMC neutralized by pulmonary surfactant should be assessed to answer the question of its applicability as a permeation enhancer in inhalation to the alveolar region.

# Acknowledgement

We thank Chloé Puisney, Milad Radiom and Nicolas Tsapis for fruitful discussions. ANR (Agence Nationale de la Recherche) and CGI (Commissariat à l'Investissement d'Avenir) are gratefully acknowledged for their financial support of this work through Labex SEAM (Science and Engineering for Advanced Materials and devices) ANR-10-LABX-0096 et ANR-18-IDEX-0001. We acknowledge the ImagoSeine facility (Jacques Monod Institute, Paris, France), and the France BioImaging infrastructure supported by the French National Research Agency (ANR-10-INBS-04, « Investments for the future »). This research was supported in part by the Agence Nationale de la Recherche under the contract ANR-13-BS08-0015 (PANORAMA), ANR-12-CHEX-0011 (PULMONANO), ANR-15-CE18-0024-01 (ICONS), ANR-17-CE09-0017 (AlveolusMimics) and by Solvay. We also want to thank for the financial support of Brno University of Technology.

# Author contributions

Jana Szabová – Conceptualization, Investigation, Data curation, Writing – Original Draft.
Filip Mravec – Supervision, Writing – Review & Editing.
Rémi Le Borgne – Investigation.
Michal Kalina – Investigation, Data curation.
Mostafa Mokhtari – Resources.
Jean– François Berret – Conceptualization, Methodology, Supervision, Data curation, Formal analysis, Writing – Original Draft, Writing – Review & Editing, Funding acquisition.

# References

[1] A. Elhissi, Liposomes for Pulmonary Drug Delivery: The Role of Formulation and Inhalation Device Design, Curr. Pharm. Des. 23 (2017) 362-372.

[2] A. Misra, K. Jinturkar, D. Patel, J. Lalani, M. Chougule, Recent advances in liposomal dry powder formulations: preparation and evaluation, Expert Opin. Drug Deliv. 6 (2009) 71-89.

[3] G. Pilcer, K. Amighi, Formulation strategy and use of excipients in pulmonary drug delivery, Int. J. Pharm. 392 (2010) 1-19.

[4] W. de Kruijf, C. Ehrhardt, Inhalation delivery of complex drugs — the next steps, Curr. Opin. Pharmacol. 36 (2017) 52-57.

[5] J.S. Patton, P.R. Byron, Inhaling medicines: Delivering drugs to the body through the lungs, Nat. Rev. Drug. Discov. 6 (2007) 67-74.

[6] B.I. Florea, M. Thanou, H.E. Junginger, G. Borchard, Enhancement of bronchial octreotide absorption by chitosan and N-trimethyl chitosan shows linear in vitro/in vivo correlation, J. Control. Release 110 (2006) 353-361.

[7] S. Maher, D. Brayden, L. Casettari, L. Illum, Application of Permeation Enhancers in Oral Delivery of Macromolecules: An Update, Pharmaceutics 11 (2019) 41.

[8] M. Sakagami, Systemic delivery of biotherapeutics through the lung: opportunities and challenges for improved lung absorption, Ther. Deliv. 4 (2013) 1511-1525.

[9] E.D. Freitas, C.F. Moura Jr, J. Kerwald, M.M. Beppu, An Overview of Current Knowledge on the Properties, Synthesis and Applications of Quaternary Chitosan Derivatives, Polymers 12 (2020) 2878.

[10] A.D. Kulkarni, H.M. Patel, S.J. Surana, Y.H. Vanjari, V.S. Belgamwar, C.V. Pardeshi, N,N,N-Trimethyl chitosan: An advanced polymer with myriad of opportunities in nanomedicine, Carbohydr. Polym. 157 (2017) 875-902.

[11] V.K. Mourya, N.N. Inamdar, Trimethyl chitosan and its applications in drug delivery, J. Mater. Sci.: Mater. Med. 20 (2009) 1057-1079.

[12] A. Jintapattanakit, S. Mao, T. Kissel, V.B. Junyaprasert, Physicochemical properties and biocompatibility of N-trimethyl chitosan: Effect of quaternization and dimethylation, Eur. J. Pharm. Biopharm. 70 (2008) 563-571.

[13] J.K. Sahni, S. Chopra, F.J. Ahmad, R.K. Khar, Potential prospects of chitosan derivative trimethyl chitosan chloride (TMC) as a polymeric absorption enhancer: synthesis, characterization and applications, J. Pharm. Pharmacol. 60 (2010) 1111-1119.

[14] D.P. Facchi, S.P. Facchi, N,N,N-Trimethyl Chitosan and Its Potential Bactericidal Activity: Current Aspects and Technological Applications, J. Infect. Dis. Ther. 4 (2016) 1000291.

[15] B.-I. Andreica, X. Cheng, L. Marin, Quaternary ammonium salts of chitosan. A critical overview on the synthesis and properties generated by quaternization, Eur. Polym. J. 139 (2020) 110016.

[16] B.E. Benediktsdóttir, Ó. Baldursson, M. Másson, Challenges in evaluation of chitosan and trimethylated chitosan (TMC) as mucosal permeation enhancers: From synthesis to in vitro application, J. Control. Release 173 (2014) 18-31.

[17] A. Fabiano, D. Beconcini, C. Migone, A.M. Piras, Y. Zambito, Quaternary Ammonium Chitosans: The Importance of the Positive Fixed Charge of the Drug Delivery Systems, Int. J. Mol. Sci. 21 (2020) 6617.

[18] O.H. Wittekindt, Tight junctions in pulmonary epithelia during lung inflammation, Pflug. Arch. Eur. J. Physiol. 469 (2017) 135-147.

[19] E. Lopez-Rodriguez, J. Perez-Gil, Structure-Function Relationships in Pulmonary Surfactant Membranes: From Biophysics to Therapy, Biochim. Biophys. Acta, Biomembr. 1838 (2014) 1568-1585.

[20] K.W. Lu, J. Pérez-Gil, M. Echaide, H.W. Taeusch, Pulmonary surfactant proteins and polymer combinations reduce surfactant inhibition by serum, Biochim. Biophys. Acta, Biomembr. 1808 (2011) 2366-2373.

[21] L.-P.-A. Thai, F. Mousseau, E. Oikonomou, M. Radiom, J.-F. Berret, Effect of Nanoparticles on the Bulk Shear Viscosity of a Lung Surfactant Fluid, ACS Nano 14 (2020) 466-475.

[22] N. Hobi, G. Siber, V. Bouzas, A. Ravasio, J. Perez-Gil, T. Haller, Physiological Variables Affecting Surface Film Formation by Native Lamellar Body-Like Pulmonary Surfactant Particles, Biochim. Biophys. Acta, Biomembr. 1838 (2014) 1842-1850.

[23] T. Kobayashi, A. Shido, K. Nitta, S. Inui, M. Ganzuka, B. Robertson, The critical concentration of surfactant in fetal lung liquid at birth, Resp. Physiol. 80 (1990) 181-192.

[24] M. Numata, P. Kandasamy, D.R. Voelker, Anionic Pulmonary Surfactant Lipid Regulation of Innate Immunity, Expert Rev. Respir. Med. 6 (2012) 243-246.

[25] A. Hidalgo, A. Cruz, J. Perez-Gil, Pulmonary Surfactant and Nanocarriers: Toxicity *versus* Combined Nanomedical Applications, Biochim. Biophys. Acta, Biomembr. 1859 (2017) 1740-1748.

[26] M. Beck-Broichsitter, Biophysical Activity of Impaired Lung Surfactant upon Exposure to Polymer Nanoparticles, Langmuir 32 (2016) 10422-10429.

[27] A. Braun, P.C. Stenger, H.E. Warriner, J.A. Zasadzinski, K.W. Lu, H.W. Taeusch, A Freeze-Fracture Transmission Electron Microscopy and Small Angle X-Ray Diffraction Study of the Effects of Albumin, Serum, and Polymers on Clinical Lung Surfactant Microstructure, Biophys. J. 93 (2007) 123-139.

[28] G.A. Georgiev, C. Vassilieff, A. Jordanova, A. Tsanova, Z. Lalchev, Foam film study of albumin inhibited lung surfactant preparations: effect of added hydrophilic polymers, Soft Matter 8 (2012) 12072-12079

[29] N. Kang, Z. Policova, G. Bankian, M.L. Hair, Y.Y. Zuo, A.W. Neumann, E.J. Acosta, Interaction between chitosan and bovine lung extract surfactants, Biochim. Biophys. Acta, Biomembr. 1778 (2008) 291-302.

[30] H.W. Taeusch, J.B. de la Serna, J. Perez-Gil, C. Alonso, J.A. Zasadzinski, Inactivation of Pulmonary Surfactant Due to Serum-Inhibited Adsorption and Reversal by Hydrophilic Polymers: Experimental, Biophys. J. 89 (2005) 1769-1779.

[31] E. Lopez-Rodriguez, A. Cruz, R.P. Richter, H.W. Taeusch, J. Pérez-Gil, Transient Exposure of Pulmonary Surfactant to Hyaluronan Promotes Structural and Compositional Transformations into a Highly Active State, J. Biol. Chem. 288 (2013) 29872-29881.

[32] K.W. Lu, J. Goerke, J.A. Clements, H.W. Taeusch, Hyaluronan Decreases Surfactant Inactivation In Vitro, Ped. Res. 57 (2005) 237-241.

[33] K.W. Lu, H.W. Taeusch, J.A. Clements, Hyaluronan with dextran added to therapeutic lung surfactants improves effectiveness in vitro and in vivo, Experimental Lung Research 39 (2013) 191-200.

[34] F. Mousseau, J.F. Berret, The role of surface charge in the interaction of nanoparticles with model pulmonary surfactants, Soft Matter 14 (2018) 5764-5774.

[35] F. Mousseau, C. Puisney, S. Mornet, R. Le Borgne, A. Vacher, M. Airiau, A. Baeza-Squiban, J.-F. Berret, Supported Pulmonary Surfactant Bilayers on Silica Nanoparticles: Formulation, Stability and Impact on Lung Epithelial Cells, Nanoscale 9 (2017) 14967-14978.

[36] M. Radiom, M. Sarkis, O. Brookes, E.K. Oikonomou, A. Baeza-Squiban, J.F. Berret, Pulmonary Surfactant Inhibition of Nanoparticle Uptake by Alveolar Epithelial Cells, Sci. Rep. 10 (2020) 19436.

[37] V. Nele, M.N. Holme, U. Kauscher, M.R. Thomas, J.J. Doutch, M.M. Stevens, Effect of Formulation Method, Lipid Composition, and PEGylation on Vesicle Lamellarity: A Small-Angle Neutron Scattering Study, Langmuir 35 (2019) 6064-6074.

[38] F. Mousseau, L. Vitorazi, L. Herrmann, S. Mornet, J.F. Berret, Polyelectrolyte assisted charge titration spectrometry: Applications to latex and oxide nanoparticles, J. Colloid Interface Sci. 475 (2016) 36-45.

[39] T. Curstedt, H.L. Halliday, C.P. Speer, A Unique Story in Neonatal Research: The Development of a Porcine Surfactant, Neonatology 107 (2015) 321-329.

[40] F. Mousseau, R. Le Borgne, E. Seyrek, J.F. Berret, Biophysicochemical Interaction of a Clinical Pulmonary Surfactant with Nanoalumina, Langmuir 31 (2015) 7346-7354.

[41] R. Ramanathan, J.J. Bhatia, K. Sekar, F.R. Ernst, Mortality in preterm infants with respiratory distress syndrome treated with poractant alfa, calfactant or beractant: a retrospective study, Journal of Perinatology 33 (2013) 119-125.

[42] A. Kuzmov, T. Minko, Nanotechnology approaches for inhalation treatment of lung diseases, J. Control. Release 219 (2015) 500-518.

[43] P. Job, Studies on the Formation of Complex Minerals in Solution and on Their Stability, Annales de Chimie 9 (1928) 113-203.

[44] J.S. Renny, L.L. Tomasevich, E.H. Tallmadge, D.B. Collum, Method of Continuous Variations: Applications of Job Plots to the Study of Molecular Associations in Organometallic Chemistry, Angew. Chem. Int. Ed. 52 (2013) 11998-12013.

[45] L.F. Ferreira, A.S. Picco, F.E. Galdino, L.J.C. Albuquerque, J.-F. Berret, M.B. Cardoso, Nanoparticle– Protein Interaction: Demystifying the Correlation between Protein Corona and Aggregation Phenomena, ACS Appl. Mater. Interfaces 14 (2022) 28559-28569.

[46] L. Belloni, Ionic Condensation and Charge Renormalization in Colloid Suspensions, Colloids Surf. A 140 (1998) 227 - 243.

[47] F. Mousseau, J.-F. Berret, E.K. Oikonomou, Design and Applications of a Fluorescent Labeling Technique for Lipid and Surfactant Preformed Vesicles, ACS Omega 4 (2019) 10485-10493.

[48] A.E. Neitzel, Y.N. Fang, B. Yu, A.M. Rumyantsev, J.J. de Pablo, M.V. Tirrell, Polyelectrolyte Complex Coacervation across a Broad Range of Charge Densities, Macromolecules 54 (2021) 6878-6890.

[49] S. Chen, Z.-G. Wang, Driving force and pathway in polyelectrolyte complex coacervation, Proc. Natl. Acad. Sci. 119 (2022) e2209975119.

[50] Q. Wang, J.B. Schlenoff, The Polyelectrolyte Complex/Coacervate Continuum, Macromolecules 47 (2014) 3108-3116.

[51] R. Asasutjarit, T. Theerachayanan, P. Kewsuwan, S. Veeranondha, A. Fuongfuchat, G.C. Ritthidej, Gamma sterilization of diclofenac sodium loaded- N-trimethyl chitosan nanoparticles for ophthalmic use, Carbohydr. Polym. 157 (2017) 603-612.

[52] P.V.A. Bueno, P.R. Souza, H.D.M. Follmann, A.G.B. Pereira, A.F. Martins, A.F. Rubira, E.C. Muniz, N,N-Dimethyl chitosan/heparin polyelectrolyte complex vehicle for efficient heparin delivery, Int. J. Bio. Macromol. 75 (2015) 186-191.

[53] L. Marci, M.C. Meloni, A.M. Maccioni, C. Sinico, F. Lai, M.C. Cardia, Formulation and characterization studies of trimethyl chitosan / sodium alginate nanoparticles for targeted drug delivery, ChemistrySelect 1 (2016) 669-674.

[54] A.F. Martins, P.V.A. Bueno, E.A.M.S. Almeida, F.H.A. Rodrigues, A.F. Rubira, E.C. Muniz, Characterization of N-trimethyl chitosan/alginate complexes and curcumin release, Int. J. Bio. Macromol. 57 (2013) 174-184.

[55] M. Baghaei, F.S.M. Tekie, M.R. Khoshayand, R. Varshochian, M. Hajiramezanali, M.J. Kachousangi, R. Dinarvand, F. Atyabi, Optimization of chitosan-based polyelectrolyte nanoparticles for gene delivery, using design of experiment: in vitro and in vivo study, Materials Science and Engineering: C 118 (2021) 111036.

[56] H. Liu, B. Ojha, C. Morris, M. Jiang, E.P. Wojcikiewicz, P.P.N. Rao, D. Du, Positively Charged Chitosan and N -Trimethyl Chitosan Inhibit Aβ40 Fibrillogenesis, Biomacromolecules 16 (2015) 2363-2373.

[57] P. Ramalingam, Y.T. Ko, Enhanced Oral Delivery of Curcumin from N-trimethyl Chitosan Surface-Modified Solid Lipid Nanoparticles: Pharmacokinetic and Brain Distribution Evaluations, Pharm. Res. 32 (2015) 389-402.

[58] D. Liao, X. Liu, W. Dai, T. Tang, G. Ou, K. Zhang, M. Han, R. Kang, S. Yang, D. Xiang, N -trimethyl chitosan (TMC)-modified microemulsions for improved oral bioavailability of puerarin: preparation and evaluation, Drug Delivery 22 (2015) 516-521.

[59] S. Moghassemi, E. Parnian, A. Hakamivala, M. Darzianiazizi, M.M. Vardanjani, S. Kashanian, B. Larijani, K. Omidfar, Uptake and transport of insulin across intestinal membrane model using trimethyl chitosan coated insulin niosomes, Materials Science and Engineering: C 46 (2015) 333-340.

[60] J. Cao, J. Sun, X. Wang, X. Li, Y. Deng, N -Trimethyl chitosan-coated multivesicular liposomes for oxymatrine oral delivery, Drug. Dev. Ind. Pharm. 35 (2009) 1339-1347.

[61] H. Chen, J. Wu, M. Sun, C. Guo, A. Yu, F. Cao, L. Zhao, Q. Tan, G. Zhai, N-trimethyl chitosan chloride-coated liposomes for the oral delivery of curcumin, Journal of Liposome Research 22 (2012) 100-109.

[62] F. Hu, Z. Zhou, Q. Xu, C. Fan, L. Wang, H. Ren, S. Xu, Q. Ji, X. Chen, A novel pH-responsive quaternary ammonium chitosan-liposome nanoparticles for periodontal treatment, Int. J. Bio. Macromol. 129 (2019) 1113-1119.

[63] A. Huang, A. Makhlof, Q. Ping, Y. Tozuka, H. Takeuchi, N-trimethyl chitosan-modified liposomes as carriers for oral delivery of salmon calcitonin, Drug Delivery 18 (2011) 562-569.

[64] A. Karewicz, D. Bielska, A. Loboda, B. Gzyl-Malcher, J. Bednar, A. Jozkowicz, J. Dulak, M. Nowakowska, Curcumin-containing liposomes stabilized by thin layers of chitosan derivatives, Colloids Surf. B 109 (2013) 307-316.

[65] W.-L. Chen, Z.-Q. Yuan, Y. Liu, S.-D. Yang, C.-G. Zhang, J.-Z. Li, W.-J. Zhu, F. Li, X.-F. Zhou, Y.-M. Lin, Liposomes coated with N-trimethyl chitosan to improve the absorption of harmine in vivo and in vitro, Int. J. Nanomed. 11 (2016) 325-336.

[66] Z. Zhou, F. Hu, S. Hu, M. Kong, C. Feng, Y. Liu, X. Cheng, Q. Ji, X. Chen, pH-Activated nanoparticles with targeting for the treatment of oral plaque biofilm, J. Mater. Chem. B 6 (2018) 586-592.

[67] J.-F. Berret, Controlling Electrostatic Co-Assembly Using Ion-Containing Copolymers: from Surfactants to Nanoparticles, Adv. Colloids Interface Sci. 167 (2011) 38-48.

[68] J. Fresnais, C. Lavelle, J.-F. Berret, Nanoparticle Aggregation Controlled by Desalting Kinetics, J. Phys. Chem. C 113 (2009) 16371-16379.

[69] D. Priftis, K. Megley, N. Laugel, M. Tirrell, Complex coacervation of poly(ethylene-imine)/polypeptide aqueous solutions: Thermodynamic and rheological characterization, J. Colloid Interface Sci. 398 (2013) 39-50.

[70] M. Schulz, A. Olubummo, W.H. Binder, Beyond the lipid-bilayer: interaction of polymers and nanoparticles with membranes, Soft Matter 8 (2012) 4849-4864.

[71] R.R. Netz, D. Andelman, Neutral and charged polymers at interfaces, Physics Reports-Review Section of Physics Letters 380 (2003) 1-95.

[72] D. Waisman, D. Danino, Z. Weintraub, J. Schmidt, Y. Talmon, Nanostructure of the Aqueous Form of Lung Surfactant of Different Species Visualized by Cryo-Transmission Electron Microscopy, Clin. Physiol. Funct. Imaging 27 (2007) 375-380.

[73] C. Schleh, C. Muhlfeld, K. Pulskamp, A. Schmiedl, M. Nassimi, H.D. Lauenstein, A. Braun, N. Krug, V.J. Erpenbeck, J.M. Hohlfeld, The Effect of Titanium Dioxide Nanoparticles on Pulmonary Surfactant Function and Ultrastructure, Respir. Res. 10 (2009) 90.

[74] J.-F. Berret, F. Mousseau, R. Le Borgne, E.K. Oikonomou, Sol-gel transition induced by alumina nanoparticles in a model pulmonary surfactant, Colloids Surf. A 646 (2022) 128974.

[75] X. Yang, Z. Sun, D. Wang, W. Forsling, Surface acid–base properties and hydration/dehydration mechanisms of aluminum (hydr)oxides, J. Colloid Interface Sci. 308 (2007) 395-404.

[76] L.-P.-A. Thai, F. Mousseau, E. Oikonomou, J.-F. Berret, On the Rheology of Pulmonary Surfactant: Effects of Concentration and Consequences for the Surfactant Replacement Therapy, Colloids Surf. B 178 (2019) 337-345.